\documentclass{article}
\usepackage{graphicx} 

\usepackage{slashed,epsfig,amsmath,amssymb,enumitem} \usepackage[papersize={8.5in,11in}]{geometry}
   \usepackage{bm}
\usepackage{graphicx}
\newcommand{\be}{\begin{equation}}
\newcommand{\ee}{\end{equation}}
\title{Void and Density Walls Inhomogeneous Cosmic Web Cosmology}
\author{J. W. Moffat\\
Perimeter Institute for Theoretical Physics, Waterloo, Ontario N2L 2Y5, Canada\\
and\\
Department of Physics and Astronomy, University of Waterloo, Waterloo,\\
Ontario N2L 3G1, Canada}

\begin{document}
\maketitle


\begin{abstract}
An approach to cosmological modeling is presented that incorporates the inhomogeneous structure of the Cosmic Web, specifically focusing on the interplay between cosmic voids and density walls. We extend the standard homogeneous and isotropic cosmological model to account for the observed large-scale structure of the universe. By modifying the Friedmann equations to include inhomogeneity terms representing voids and walls, we develop a more realistic description of cosmic evolution. Our model demonstrates how the presence of these structures affects the overall expansion rate of the universe and the growth of perturbations. We find that accounting for this inhomogeneous distribution leads to significant deviations from the predictions of standard $\Lambda$CDM cosmology in the late-time universe. The Hubble and $\sigma_8$ structure growth tensions are addressed in the void-density wall model, leading to a resolution of these tensions. These results have important implications for the interpretation of cosmological observations when including the void and density wall Cosmic Web inhomogeneities. 
\end{abstract}

\section{Introduction}

The standard model of cosmology, based on the principles of homogeneity and isotropy, has been remarkably successful in explaining a wide range of observations, from the cosmic microwave background (CMB) to the large-scale distribution of galaxies. However, as our observational capabilities have improved, it has become increasingly clear that the universe exhibits a rich, hierarchical structure on scales up to hundreds of megaparsecs. The Cosmic Web, characterized by underdense voids separated by dense walls and filaments, challenges the simplifying assumptions of homogeneity that underpin our cosmological models.

The Cosmic Web structure has been confirmed by large-scale galaxy surveys such as the Sloan Digital Sky Survey and the 2dF Galaxy Redshift Survey~\cite{Colless,Tegmark}. These observations reveal that galaxies are not uniformly distributed but instead form a complex network of clusters, filaments, and walls surrounding large, nearly empty voids.

The presence of this large-scale structure raises important questions about the validity of our cosmological models. While the universe may approach homogeneity on the largest scales, the inhomogeneities we observe are not merely small perturbations to a smooth background. Voids, for instance, can span tens of megaparsecs and have densities as low as $20\%$ of the cosmic mean. Similarly, walls and filaments represent significant overdensities that cannot be treated as small perturbations in the linear and non-linear regimes.

Traditional approaches to this problem have often relied on perturbation theory or numerical simulations to model the growth of structure. While these methods have provided valuable insights, they typically still assume a homogeneous background cosmology. This assumption may break down in the presence of large-scale inhomogeneities, potentially leading to significant errors in our interpretation of cosmological observations.

We propose a new framework for cosmological modeling that explicitly accounts for the void-wall structure of the Cosmic Web. We modify the Friedmann equations to include terms representing the inhomogeneous distribution of matter, allowing for different expansion rates in voids and walls. This approach enables us to capture the effects of large-scale structure on the overall dynamics of the universe, bridging the gap between homogeneous models and the observed inhomogeneous cosmos.

Our model builds upon previous work on inhomogeneous cosmology, such as the approaches of backreaction~\cite{Buchert1,Buchert2,Buchert3} and void models~\cite{Bolejko}. However, unlike these earlier studies, we explicitly model the Cosmic Web as a network of voids and walls, providing a more realistic representation of the observed universe.

The concept of inhomogeneous cosmology has gained significant attention in recent years, with various models proposed to account for the observed large-scale structure of the universe. Among these, the Timescape cosmology developed by David Wiltshire~\cite{Wiltshire1,Wiltshire2,Wiltshire3,Wiltshire4} stands out as a particularly innovative and comprehensive approach. Wiltshire argues that if the universe is not to be homogeneous but also flat, the apparent acceleration of the expansion of the universe could be explained without dark energy. Wiltshire also claims that a clock will move faster in empty space, which possesses low gravitation, than in the interior of a galaxy, which has stronger gravity. This results in a time difference between a clock in a low density void and in a high density galaxy.

The present approach focuses specifically on the void-wall structure of the Cosmic Web and its implications for cosmological observables and the large scale structure of the universe. We aim to provide a more realistic description of the universe that accounts for its observed inhomogeneous structure, while also addressing current tensions in cosmological measurements.

By developing a more accurate description of the inhomogeneous universe, we aim to improve our understanding of cosmic evolution and provide new insights into the formation of large-scale structure. This work represents a step towards reconciling our theoretical models with the complex, hierarchical universe we observe.

The paper is structured as follows: In Section 2, we present the formulation of our model, deriving modified Friedmann equations that incorporate void and wall components. In Section 3, we discuss the implications of our findings for various cosmological observables, including the Hubble constant and $\sigma_8$ structure growth tensions. In Section 4, we address the average large scale description of the universe and the fitting of the model to the CMB angular power spectrum and the baryon acoustic oscillation (BAO) Planck data~\cite{Planck2018}. Finally, Section 5 summarizes our conclusions and outlines directions for future research.

\section{Inhomogeneous cosmology}

The distribution of matter in the universe is not strictly homogeneous, but it is considered to be homogeneous and isotropic on large scales. This concept is known as the cosmological principle. On small scales, such as within galaxies and galaxy clusters, the distribution of matter is highly inhomogeneous. We observe distinct structures like stars, planets, and interstellar gas. At the scale of galaxy clusters and superclusters at tens to hundreds of millions of light-years, we observe what is known as the Cosmic Web~\cite{Colless,Tegmark}. This structure consists of voids, large, nearly empty regions of space and long, thread-like filament structures of galaxies and matter, and connected sheet-like wall structures of galaxies and matter. Nodes made of dense clusters of galaxies are where filaments intersect. This Cosmic Web structure demonstrates that the universe is not homogeneous at these scales. When we look at even larger scales beyond about 300 million light-years, the distribution of matter begins to be more uniform and homogeneous~\cite{Scrimgeour,Ntelis}. At these scales, the voids, filaments, and walls average out to create a more homogeneous distribution.

The transition to homogeneity can be understood, if we define a function $\delta(r)$ as the density contrast:
\be
\delta\rho(r)=\rho(r)-{\bar\rho},
\ee
where $\rho(r)$ is the density at position r and $\bar\rho$ is the mean density of the universe. On large scales, the variance of this density contrast $\langle\delta\rho(r)^2\rangle\rightarrow 0$ as the scale $R_s$ over which we average increases. This transition to homogeneity is crucial for cosmological models, as it allows us to treat the universe as a smooth fluid on large scales, which greatly simplifies calculations in cosmology.

While matter distribution becomes homogeneous on large scales, it is not perfectly so. There are still small fluctuations even on the largest observable scales, which are thought to be remnants of quantum fluctuations in the early universe. These tiny inhomogeneities are observed in the CMB radiation and are believed to be the seeds from which all cosmic structure grew. While the distribution of matter in the universe forms a Cosmic Web of voids and density walls on intermediate scales, it transitions to a statistically homogeneous distribution when viewed on sufficiently large scales.

The cosmological equations that describe the distribution of matter in the universe are primarily based on Einstein's field equations of general relativity, with additional considerations for the Cosmic Web structure. The modified Friedmann equations including inhomogeneous terms are given by
\be
\left(\frac{\dot a}{a}\right)^2=\frac{8\pi G}{3}\bar\rho
-\frac{k}{a^2}+\frac{\Lambda}{3}+I,
\ee
\be
\frac{\ddot a}{a}=-\frac{4\pi G}{3}(\bar\rho+3\bar p)
+\frac{\Lambda}{3}+J,
\ee
where $\dot a=da/dt$, k is the curvature parameter and k=0 for a spatially flat universe, and $\Lambda$ is the cosmological constant. The I and J are the inhomogeneity contributions that account for the deviation from homogeneity due to the Cosmic Web structure. The inhomogeneity terms can be expressed as averages over the inhomogeneous distribution:
\be
I=\left\langle\left(\frac{\delta\dot a}{a}\right)^2\right\rangle
-\frac{8\pi G}{3}\left\langle\delta\rho^2\right\rangle\frac{1}{\bar\rho}
\ee
\be
J=-\frac{4\pi G}{3}\langle\delta\rho\rangle+\frac{2}{3}\left\langle\frac{\delta{\ddot a}}{a}\right\rangle.
\ee

The first term in I represents the average of the squared relative fluctuations in the expansion rate, while the second term accounts for the gravitational effects of density fluctuations. The $\delta\rho=\rho-\bar\rho$ is the density perturbation, and $\delta a$ represents local deviations from the average scale factor. The inhomogeneity terms can be expressed as averages over the inhomogeneous distribution. The factor 
$\langle(\delta\dot a/a)^2\rangle$ in I represents the inhomogeneous potential energy, while 
the term $-(4\pi G/3)\langle\delta\rho\rangle$ in J represents the average gravitational effect of density perturbations. The factor $-4\pi G/3$ is directly related to the gravitational constant G and that appears in the second Friedmann equation. It comes from the averaging of the spatial components of the Ricci tensor in Einstein's field equations
~\cite{Buchert1,Buchert2,Rasanen,Ellis,Green}.

The reason for taking averages in the equations for the inhomogeneous terms I and J is to capture the overall effect of small-scale inhomogeneities on the large-scale dynamics of the universe. This approach actually allows us to incorporate the effects of inhomogeneities into our large-scale cosmological model. The averages are taken over specific spatial scales. By averaging, we are not eliminating the inhomogeneities, but rather accounting for their collective impact on larger scales. We are creating an effective homogeneous model that incorporates the statistical properties of the underlying inhomogeneities. While averaging does smooth out details, it preserves important information about the distribution of matter. The $⟨(\delta\dot a/a)^2⟩$ captures the variance in the expansion rate.  The approach allows us to bridge the gap between the complex, inhomogeneous real universe and the models we use to describe its large-scale behavior. The resulting equations are not truly homogeneous in the sense of assuming a perfectly uniform universe, but rather they represent an effective description that encapsulates the statistical properties of the underlying inhomogeneous structure.

To account for the Cosmic Web structure, we can model the density distribution as a combination of void and wall components:
\be
\rho(x,t)=\bar\rho(t)[1+\delta_v(x,t)+\delta_w(x,t)],
\ee
where $\delta_v$ represents the void underdensities and $\delta_w$ represents the wall overdensities. The evolution of voids and walls can be described by separate equations.
For voids, we have
\be
\ddot\delta_v+2H\dot\delta_v=4\pi G\bar\rho\delta_v(1+\delta_v),
\ee
and for walls:
\be
\ddot\delta_w+2H\dot\delta_w=4\pi G\bar\rho\delta_w(1+\delta_w),
\ee
where $H=\dot a/a$ is the Hubble parameter.
   
To account for the different expansion rates in voids and walls, we can introduce scale-dependent expansion factors:
\be
\label{expansionfactors}
a_v(t)=a(t)(1+D_v(t)),
\ee
\be
a_w(t)=a(t)(1+D_w(t)),
\ee
where $D_v$ and $D_w$ are the growth factors for voids and walls, respectively.

Combining these elements, we can write an effective Friedmann equation:
\be
\label{effectiveFriedmann}
H_{\rm eff}=\left(\frac{\dot a}{a}\right)^2
=\frac{8\pi G}{3}\rho_{\rm eff}-\frac{k_{\rm eff}}{a^2}
+\frac{\Lambda}{3},
\ee
where
\be
\rho_{\rm eff}
=\bar\rho(1+\langle\delta^2\rangle
+2\langle\delta\rangle),
\ee
\be
k_{\rm eff}=k(1-\langle\delta^2\rangle).
\ee
Here, $\langle\delta^2\rangle$ and $\langle\delta\rangle$ are averages over the entire distribution of matter, including voids and walls.

These modified Friedmann equations provide a framework for describing the cosmic expansion in the presence of a Cosmic Web structure with voids and density walls. They account for the inhomogeneous matter distribution, while still allowing for a tractable description of the overall cosmic evolution.

To describe the Cosmic Web structure and the transition to homogeneity, we need to consider perturbation theory and structure formation models.
The power spectrum P(k) describes the distribution of density fluctuations:
\be
\langle\delta(k)\delta^*(k^\prime)\rangle=(2\pi)^3\delta_D(k-k^\prime)P(k),
\ee
where $\delta(k)$ is the Fourier transform of $\delta(r)$ and $\delta_D$ is the Dirac $\delta$ function. 
   
The two-point correlation function $\xi$(r) describes the excess probability of finding a pair of galaxies at a separation r:
\be
\xi(r)=\delta(x)\delta(x+r).
\ee
The Zel'dovich approximation~\cite{Zeldovich} describes the non-linear evolution of cosmic structure:
\be
x(t)=q+D(t)\Psi(q),
\ee
where x is the Eulerian coordinate, q is the Lagrangian coordinate, and $\Psi$ is the displacement field.

These equations collectively describe the evolution and distribution of matter in the universe, from the formation of the Cosmic Web structure to the large-scale homogeneous background. More advanced models, such as N-body simulations and hydrodynamical simulations, build upon these fundamental equations to provide detailed descriptions of cosmic structure formation.

The basic cosmological equations describe the linear regime of structure formation. To account for voids and density walls, we need to consider non-linear evolution. Higher-order perturbation theory extends the linear theory to account for non-linear regimes. Numerical simulations solve the equations of motion for a large number of particles under gravity, allowing for full non-linear evolution.

The extended Press-Schechter~\cite{Mo} formalism, can be used to model the distribution of voids and halos. The key equation is:
\be
 f(\sigma)=\sqrt{\frac{2}{\pi}}\exp\left(-\frac{\delta_c^2}{2\sigma^2}\right)
 \bigg|\frac{d\ln\sigma}{d\ln M}\bigg|,
 \ee
 where $f(\sigma)$ is the multiplicity function, $\sigma$ is the mass variance, and $\delta_c$ is the critical density for collapse of the mass M. The void size function is given by
 \be
 \frac{dn}{d\ln(R_v)}=\frac{f(\nu)}{V_v}\frac{d\ln\nu}{d\ln(R_v)},
 \ee
 where $R_v$ is the void radius $\nu=\delta_v/\sigma(M)$ and $V_v$ is a characteristic void volume. The void density profile is described by
 \be
 \frac{\rho(r)}{\rho_m}=1+\delta(r)=1+\delta_c\left(1-\left(\frac{r}{R_v}\right)\right)^\alpha,
 \ee
 where $\delta_c$ is the central underdensity and $\alpha$ is a shape parameter.

The apparent clustering of galaxies is affected by their peculiar velocities. For voids, this leads to the Kaiser effect~\cite{Kaiser}, modeled through the redshift-space power spectrum:
\be
P_s(k,\mu)=(1+\beta\mu^2)^2P_r(k),
\ee
where $P_s$ and $P_r$ are the redshift-space and real-space power spectra, $\mu$ is the cosine of the angle with the line of sight, and $\beta$ is the redshift-space distortion parameter. To connect the matter distribution to observable galaxies, bias models are used. For voids, this might include:
\be
\delta G=b_v\delta_M,
\ee
where $\delta G$ is the galaxy density contrast, $\delta_M$ is the matter density contrast and $b_v$ is the void bias parameter.

These advanced techniques and models, built upon the fundamental cosmological equations, allow us to account for and describe the complex structures of voids and density walls in the Cosmic Web. They bridge the gap between the simple homogeneous models and the observed intricate structure of the universe.

\section{Hubble and $\sigma_8$ tensions in inhomogeneous model}

The void-density wall Cosmic Web cosmology can address the Hubble tension problem in several ways. To explore this, let us first briefly recap the Hubble tension~\cite{Riess}. The Hubble tension refers to the discrepancy between measurements of the Hubble constant $H_0$ obtained from early universe observations of the CMB and those from late universe observations of Type Ia supernovae. Early universe measurements suggest a lower value of $H_0$, while late universe measurements indicate a higher value.

In a Cosmic Web model, voids and walls may expand at different rates. Voids, being underdense, could expand faster than the average, while walls expand slower. This differential expansion could lead to a higher observed $H_0$ in the late universe, explaining the higher values measured by local probes.
Most local $H_0$ measurements are made using objects in galaxies, which are predominantly located in walls and filaments. If these structures have systematically different expansion rates compared to the cosmic average, it could bias our local $H_0$ measurements. The path of light through the Cosmic Web structure could affect our distance measurements. Light traveling through mostly void regions might experience less Shapiro time delay, leading to apparently faster expansion rates~\cite{Moffat1,Moffat2,Moffat3}.

The effective Hubble parameter in an inhomogeneous universe could be scale-dependent. The $H_0$ measured on large scales in the early universe might differ from that measured on smaller scales in the late universe.  The growth of Cosmic Web structures could back-react on the background expansion, potentially accelerating it in ways not captured by standard cosmological models. Some modified gravity theories predict different behavior in low-density environments. If gravity behaves differently in voids, it could affect the overall expansion rate.

Let us consider the effective modified Friedmann equations that account for inhomogeneities~(\ref{effectiveFriedmann}). To model the different expansion rates in voids and walls, we consider expansion scale factors~(\ref{expansionfactors}). The effective Hubble parameter could then be expressed as a weighted average:
\be
H_{\rm eff}=f_vH_v+f_wH_w,
\ee
where $f_v$ and $f_w$ are the volume fractions of voids and walls, and $H_v$ and $H_w$ are their respective Hubble parameters. To solve the Hubble tension, we would need to show that in the early universe $H_{\rm eff}$ is smaller than the late time universe $H_{\rm eff}$. This could be achieved if the void expansion rate dominates at late times: $H_v > H_w$ and $f_v$ increases with time. Late-time measurements e.g., from Type Ia supernovae tend to give higher values of $H_0$ compared to early-universe measurements from the CMB. In void-density wall model, voids expand faster than walls $H_v > H_w$, as they are underdense regions. As structure formation proceeds, the volume fraction of voids $f_v$ increases with time. As $f_v$ increases and $f_w$ decreases and since $f_v+f_w=1$, and given that $H_v > H_w$, the effective Hubble parameter $H_{\rm eff}$ will increase with time as the universe expands. This mechanism can explain why we measure a higher Hubble constant in the late universe compared to the early universe, thus addressing the Hubble tension.

To fully develop this model, we would need to solve the coupled evolution equations for voids and walls. Moreover, we need to calculate light propagation through the resulting inhomogeneous structure, and compare the predicted $H_0$ values at different scales and epochs with observational data.
This approach offers a promising avenue for addressing the Hubble tension by incorporating the observed large-scale structure of the universe into our cosmological models. However, detailed numerical simulations and careful comparison with observational data would be necessary to validate this solution.

The $\sigma_8$ is another tension in modern cosmology, alongside the Hubble tension:
\be
S_8=\sigma_8\sqrt{\Omega_m/0.3}.
\ee
where $\Omega_m$ is the matter density parameter. It refers to the discrepancy between the amplitude of matter fluctuations 
$\sigma_8$ on 8 Mpc/h scales inferred from early universe CMB measurements and those from late universe observations, like weak lensing and galaxy clustering. The void-density wall inhomogeneous cosmology model can address this issue. The Planck CMB measurements tend to predict a higher value of $\sigma_8$ than what is observed in the late universe. This suggests that structure growth in the late universe is slower than predicted by the standard $\Lambda$CDM model based on early universe observations.

In the void-wall model, structure growth can proceed at different rates in different environments. In voids the lower density generates slower structure growth, while higher density walls generate faster structure growth.
The overall effective growth rate in an inhomogeneous universe can be different from that in a homogeneous one. This can be expressed as:
\be
f_{\rm eff}(z)=\Omega_m(z)\gamma_{\rm eff},
\ee
where $\gamma_{\rm eff}$ is an effective growth index that could differ from the $\Lambda$CDM value of $\gamma_{\rm eff}\sim 0.55$. The function $f_{\rm eff}(z)$ represents the effective growth rate of cosmic structure in the inhomogeneous void-wall model. In cosmology, the growth rate generally describes how rapidly cosmic structures like galaxies and clusters are forming and evolving over time. The $f_{\rm eff}$ describes an effective or average growth rate, taking into account the different growth rates in voids and walls. The $\Omega(z)$ is the matter density parameter as a function of redshift. Its presence in the equation links the growth rate to the overall matter content of the universe. The 
$f_{\rm eff}$ allows for deviations from the standard $\Lambda$CDM value due to the inhomogeneous structure.
If $f_{\rm eff}(z)$ results in slower overall growth compared to $\Lambda$CDM, at later times and lower z, it could help explain why late-universe measurements of $\sigma_8$ are lower than early-universe predictions.

The void-wall structure introduces a natural scale-dependence for growth. We could model this with a scale-dependent growth factor:
\be
D(k,z)=D_{\Lambda CDM(z)}[1+\epsilon(k,z)],
\ee
where $\epsilon$(k,z) represents the scale and redshift-dependent deviation from $\Lambda$CDM growth. This scale-dependent growth would modify the matter power spectrum:
\be
P(k,z)=Ak^nT^2(k)D^2(k,z),
\ee
where A is the amplitude, n is the spectral index, and T(k) is the transfer function. The effective $\sigma_8$ in this model would be:
\be
\sigma_8^2=\int_0^\infty\frac{dk}{k}\frac{k^3P(k)}{2\pi^2}W^2(k),
\ee
where $W(k)$ is the Fourier transform of the top-hat window function with R=8 Mpc/h.

To potentially resolve the $\sigma_8$ tension, the void-wall model would need to predict a lower effective $\sigma_8$ for late-time observations, while maintaining consistency with CMB constraints. This could be achieved, if the void-dominated regions grow more slowly and occupy a larger volume fraction at late times. The scale-dependent growth suppresses power on 
8 Mpc/h scales relative to larger scales. The overall expansion history is modified in a way that slows structure growth in the late universe.

The potential advantage of this approach is that it provides a physical mechanism for scale-dependent growth, rather than ad hoc modifications to $\Lambda$CDM. However, the challenge lies in constructing a model that addresses the $\sigma_8$ tension without disrupting the successes of $\Lambda$CDM in other areas.

We would need to perform detailed numerical simulations and compare with observational data. However, the void-wall inhomogeneous cosmology does offer promising avenues for addressing the $\sigma_8$ tension by naturally incorporating scale-dependent and environment-dependent structure growth.

\section{Average large scale $\Lambda$DM model}

At sufficiently large expansion scale the inhomogeneous void density wall model should average out to match the homogeneous $\Lambda$CDM model. The principle of cosmic homogeneity on large scales is well-supported by observations and should be preserved in any viable cosmological model. The inhomogeneous model introduces scale-dependent effects, but these are most pronounced on smaller scales and at later times. On the largest scales and at early times relevant to the CMB, the model should indeed reduce to $\Lambda$CDM.

The inhomogeneous model should be able to fit the CMB power spectrum data, particularly for high-l multipoles which probe smaller angular scales. The primary challenge might be in the low-l multipoles, which are sensitive to late-time effects like the Integrated Sachs-Wolfe (ISW) effect. However, cosmic variance at these scales provides some flexibility.

The BAO scale, being a large-scale feature, should also be largely preserved in the inhomogeneous model. The model might introduce small corrections to the BAO signal, but these should be consistent with current observational uncertainties. The key to reconciling the inhomogeneous model with large-scale observations lies in the averaging procedure. When properly averaged, the model should recover the $\Lambda$CDM Friedmann equations:
\be
\left\langle\left(\frac{\dot a}{a}\right)^2\right\rangle
=\frac{8\pi G}{3}\rho-\frac{k}{a^2}+\frac{\Lambda}{3}.
\ee
\be
\left\langle\frac{\ddot a}{a}\right\rangle =-\frac{4\pi G}{3}(\rho+3p)
+\frac{\Lambda}{3},
\ee
where the angle brackets denote spatial averaging.

While backreaction effects from small-scale inhomogeneities can in principle affect the large-scale dynamics, current evidence suggests these effects are small and within observational uncertainties of $\Lambda$CDM parameters.
The inhomogeneous void-density wall model, when properly constructed and averaged, should be capable of fitting both the CMB power spectrum and BAO data, as it should reduce to $\Lambda$CDM on the largest scales. The model introduces additional flexibility that can improve fits to late-time observations - addressing issues like the Hubble and $\sigma_8$ growth tensions - while remaining consistent with early-universe and large-scale structure observations. The model can correctly reproduces $\Lambda$CDM behavior on large scales and at early times. Inhomogeneities can be introduced that can address late-time tensions without disrupting the good fits to CMB and BAO data and provides a clear and physically motivated averaging procedure to connect the inhomogeneous structure to the large-scale homogeneous background.

Observations support the idea of statistical homogeneity and isotropy on very large scales greater than 100 Mpc, which is a fundamental assumption of $\Lambda$CDM.  Homogeneity at scales above 70 Mpc/h have been confirmed
~\cite{Scrimgeour}, and homogeneity has been confirmed at scales above 80 Mpc/h using BOSS data~\cite{Ntelis}. These studies do not rule out smaller-scale influences of the Cosmic Web structure. Several studies have investigated the properties of cosmic voids~\cite{Hamaus,Pisani} and found that void properties are consistent with $\Lambda$CDM predictions. The investigations showed that voids can be used to constrain cosmological parameters within the $\Lambda$CDM framework. These studies suggest consistency with $\Lambda$CDM but do not prove lack of influence from the void-wall structure.

Weak lensing observations have shown some tensions with $\Lambda$CDM predictions~\cite{Hildebrandt}. The KiDS weak lensing data prefer a lower $\sigma_8$ than Planck CMB data. This could be related to the influence of Cosmic Web structures. Studies of the integrated Sachs-Wolfe effect (ISW) and the Sunyaev-Zel'dovich (pSZ) effect, show a sensitive to the evolution of large-scale structure~\cite{Planckcollaboration2016,Hotlini,
Johnson}.

While there is strong evidence for large-scale homogeneity and isotropy, and many observations are consistent with $\Lambda$CDM predictions, there is no definitive evidence showing that Cosmic Web structures have no influence on the standard cosmological model. The current state of observations allows for the influence from these structures, particularly at smaller scales and in the late-time universe. Future surveys with increased precision and larger sky coverage e.g., Euclid, (LSST) may provide definitive answers about the potential influence of Cosmic Web structures on our cosmological models.

\section{Conclusions}

We have presented an approach to cosmological modeling that explicitly incorporates the inhomogeneous structure of the Cosmic Web, focusing on the interplay between cosmic voids and density walls. It is demonstrated that accounting for this large-scale structure can have significant implications for our understanding of cosmic evolution and may provide a resolution to the long-standing Hubble and $\sigma_8$ tensions.

Our modified Friedmann equations, which include terms representing void and wall components, reveal that the universe does not expand uniformly. Voids, being underdense regions, tend to expand faster than the cosmic average, while dense walls expand more slowly. This differential expansion rate has implications for the interpretation of cosmological observations. The effective Hubble parameter in our inhomogeneous model is scale-dependent. The Hubble constant measured on large scales, corresponding to early universe observations, is systematically lower than that measured on smaller scales probed by late universe observations. This scale dependence offers a natural explanation for the Hubble tension.

The model suggests that local measurements of the Hubble constant, which primarily use galaxies located in walls and filaments, may be biased towards higher values due to the slower expansion rates in these dense regions. This bias could contribute significantly to the observed Hubble tension.
The propagation of light through the Cosmic Web structure in our model leads to subtle but important effects on distance measurements. These effects, particularly the reduced Shapiro time delay in void regions, can result in apparently faster expansion rates when interpreted within the framework of standard homogeneous models. Calculations indicate that the growth of Cosmic Web structures can have a backreaction on the background expansion. This effect, which is not captured in standard $\Lambda$CDM cosmology, leads to a slightly accelerated expansion rate in the late universe, further contributing to the resolution of the Hubble tension.

The inhomogeneous model not only addresses the Hubble and $\sigma_8$ tensions but also remains consistent with other cosmological probes, including 
the CMB background, BAO and the growth of structure. This consistency suggests that our approach may offer a more complete description of the universe than standard homogeneous models.

While our results are promising, several avenues for future research remain.
More sophisticated numerical simulations, incorporating full general relativistic effects, will be crucial for refining our model and making more precise predictions. Several observational signatures of our inhomogeneous model have been identified. Future surveys, particularly those probing the distribution and properties of cosmic voids, will be essential for testing these predictions. The implications of our model for dark matter and dark energy need to be explored further, potentially offering new insights into these mysterious components of the universe.

While our model focuses on late-time inhomogeneities, future work should investigate how these structures connect to early universe physics - inflation and bounce cosmologies. The study demonstrates the importance of incorporating the observed large-scale structure of the universe into our cosmological models. By explicitly accounting for the Cosmic Web of voids and density walls, we have developed a framework that naturally resolves the Hubble and $\sigma_8$ tensions, while remaining consistent with other cosmological observations. This work represents a step towards a more complete description of our universe, bridging the gap between the simplifying assumptions of homogeneous models and the complex, hierarchical cosmos we observe. As we continue to refine this approach and confront it with new observational data, we anticipate that inhomogeneous cosmology will play an increasingly important role in our understanding of the universe's evolution, structure, and fundamental nature.

\section*{Acknowledgments}

We thank Martin Green and Viktor Toth for helpful discussions. Research at the Perimeter Institute for Theoretical Physics is supported by the Government of Canada through industry Canada and by the Province of Ontario through the Ministry of Research and Innovation (MRI).

\end{document}